# Photon plausibel bei linearer Polarisation [22]

Petra Schulz
Theodor-Francke-Weg 65, 38116 Braunschweig, Deutschland

**Abstract**
A plausible photon model is presented. The model is based on the idea that photon and particle must have the same structure. The linear polarization is interpreted under the assumption that vibration photons interact with the particles in the polarizer/analyzer. Slantwise flying photons are separated by the particles of the dichroic or double-refracting material into two components because a vibration component is absorbed and the other one is let through inevitably. The vibrations are explained by example of the carbonate ion of calcite.

**Kurzfassung**
Es wird ein plausibles Photonenmodell vorgelegt. Das Modell basiert auf der Vorstellung, dass Photon und Teilchen die gleiche Struktur haben müssen. Lineare Polarisation wird gedeutet unter der Annahme, dass von den Teilchen im Polarisator/Analysator Schwingungsphotonen aufgenommen werden. Bei schrägem Einschuss der Photonen werden sie von den Teilchen des dichroitischen oder doppelbrechenden Materials in zwei Komponenten getrennt, weil eine Schwingungskomponente absorbiert wird und die andere zwangsläufig durchgelassen wird. Die Schwingungen werden am Beispiel des Carbonat-Ions von Calcit erläutert.

## 1. Einleitung

Versuche zum Thema lineare Polarisation lassen sich sehr gut demonstrieren, aber nur schlecht anschaulich verstehen. Ich möchte geringfügig andere Erklärungen benutzen, als sie in den Schulbüchern zu finden sind (auch nicht in dem sehr soliden speziellen Buch über Optik von HECHT [3]). Dabei möchte ich auf die Begriffe elektrisches Feld und magnetisches Feld verzichten, weil sie zu abstrakt sind und im Grunde genommen gar nicht die tragende Rolle spielen. Das elektromagnetische Strahlungsfeld wird durch Photonen ersetzt, die passgerecht auf die Teilchen im Polarisator zugeschnitten sind.

Als Polarisator-/Analysatormaterial beschränke ich mich auf doppelbrechende und dichroitische Kristalle. Beide Festkörper haben eine gemeinsame Eigenschaft: Längs der Kristallachsen beobachtet man unterschiedliche Brechzahlen. Diesen Sachverhalt in die Sprache der Spektroskopie übersetzt: Für unterschiedliche Schwingungsrichtungen benötigt der Kristall unterschiedliche Absorptionsfrequenzen.

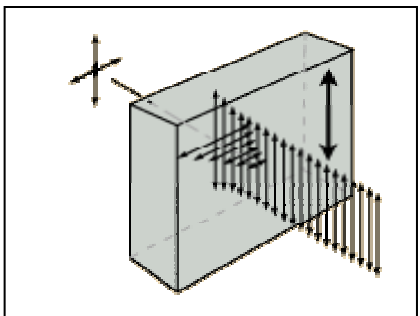

**Abb. 1**: Beispiel für einen dichroitischen Kristall: Die senkrecht schwingende Lichtkomponente wird durchgelassen, die waagerechte absorbiert [nach 21]

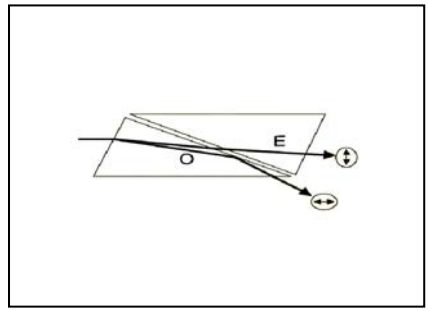

**Abb. 2**: Doppelbrechung im Nicolschen Prisma: Der senkrecht schwingende Strahl (E) wird durchgelassen und der waagerecht schwingende (O) total reflektiert und an der Polarisator-/Analysator-Wandung absorbiert

Dichroitische Kristalle absorbieren eine Schwingungsrichtung des einfallenden Lichts (Abb. 1). Bei doppelbrechenden Kristallen gestaltet sich der Polarisationseffekt umständlicher. Schickt man z. B. auf die zusammengeklebten Prismen eines Nicolschen Prismas einen Lichtstrahl schräg zur optischen Achse, so spaltet er in Teilstrahlen auf. Ein Strahl einer Polarisationsrichtung tritt aus, der andere mit der anderen Polarisationsrichtung wird totalreflektiert und von der geschwärzten Polarisatorwandung absorbiert (Abb. 2). Beim Nicolschen Prisma ereignet sich also eine indirekte Absorption.

Im folgenden werden Kapitel, die spezielle Vorkenntnisse des Lesers verlangen, mit einem „ * " gekennzeichnet, und dürfen übersprungen werden. Der ganz eilige Leser (Physikdidaktiker) kann sofort zum Kapitel 2.2.5 übergehen, der halbeilige Leser zu Kapitel 2.2.3, wo das Thema Photon gestreift wird, oder der mitteleilige Leser zu 2.2.

## 2. Vorbemerkungen - Hintergrundwissen
### 2.1 Zur Technik und zur Taktik *

Die stabilen robusteren Polarisatoren bestehen aus (ein)kristallinen Festkörpern (und die weniger robusten aus festkörperähnlichen Stoffen – aus kristallinen Flüssigkeiten, die auf einer Folie aufgezogen sind). In der Schule werden die Polarisationsversuche im sichtbaren Bereich durchgeführt. In der Durchlassrichtung absorbieren die Filter nur geringfügig. Es ist anzunehmen,

- dass die Festkörperteilchen hauptsächlich Schwingungen ausüben.
- Im sichtbaren Bereich sind gewiss schon diverse Ober- und Kombinations- (und andere) Schwingungen angeregt. Eine genauere Kenntnis dieser Schwingungsanregung ist für ein allgemeines Verständnis wie in der Schule nicht erforderlich. Außerdem ist darüber im Moment wohl auch nichts genaues bekannt.
- Laut Lehrbuch von HELLWEGE ([4], S. 106) sind diese Schwingungen bereits kohärent, wie aus IR-Absorptionsspektren zu schließen ist.
- Ferner ist anzunehmen, dass unter gewissen äußeren Gegebenheiten (1. bei genügender Energie und 2. bei schräger Aufprallrichtung der Photonen relativ zu den schwingenden Teilchen) die in den Schwingern bereits eingefangenen Photonen, die als Vektoren betrachtet werden können, genau wie Vektoren zerlegt – sprich zerrupft - werden können. Darauf beruhen schließlich die optisch-parametrischen Oszillatoren (OPOs), die in der heutigen Lasertechnologie eine wichtige Rolle spielen (s. z. B. SCHILLER / MEYN [6]). Die Schwinger der OPOs führen vermutlich eine Kombinations-Schwingung aus einer senkrechten und einer waagerechten linearen Schwingung aus. Diese beiden Teilschwingungen werden auseinander getrennt.

Auch höhere Oberschwingungen lassen sich bei schrägem Lichteinfall je nach Einfallswinkel in Schwingungen mit halber, viertel und achtel Frequenz zerhacken (KING / DEKER 1983 [5]; Originalarbeit WEISS / KING 1982 [11]; WEISS und Mitarbeiter 1983 [12] sowie das Buch von WEISS und VILASECA 1991 [13]). Dieser Sachverhalt wurde von der Arbeitsgruppe um WEISS anders gedeutet. Auf meine Interpretationsweise werde ich in einem späteren Artikel näher eingehen.

### 2.2 Zur Theorie: Schwingungen und Photonen
#### 2.2.1 Schwingungen und Struktur

Calcit ($CaCO_3$) ist ein wichtiges Material in der Optik (es dient z. B. zur Herstellung vom Nicolschen Prisma oder ähnlichen Polarisatoren modernerer Bauweise). Calcit ist ein Musterbeispiel für einen doppelbrechenden Stoff. Ich wiederhole vorsichtshalber noch einmal: Doppelbrechend bedeutet, ein Stoff besitzt längs seiner Kristallachsen zwei unterschiedliche Brechzahlen. Ohne viele Grundkenntnisse in Kristallphysik oder –chemie zu besitzen, kann man von Calcit sehr viel lernen, was zum Thema Polarisation passt.

In Abb. 3 ist ein Stück Kristall von Calcit dargestellt. Man kann folgendes daraus entnehmen: An den Ecken und auf den Flächenmitten des Rhomboeders sitzen die Calcium-Ionen ($Ca^{2+}$), auf den Kantenmitten und in der Mitte des Rhomboeders auf der optischen Achse befinden sich die Carbonat-Ionen ($CO_3^{2-}$). Die Carbonat-Ionen sind in parallelen Schichten angeordnet. Entscheidend sind jedoch nur die Einheiten des Carbonat-Ions ($CO_3^{2-}$), das die Struktur eines gleichseitigen Dreiecks hat, in dessen Mitte ein Kohlenstoffatom sitzt und an den Ecken jeweils ein Sauerstoffatom. Das Carbonat-Ion gleicht einem Propeller.

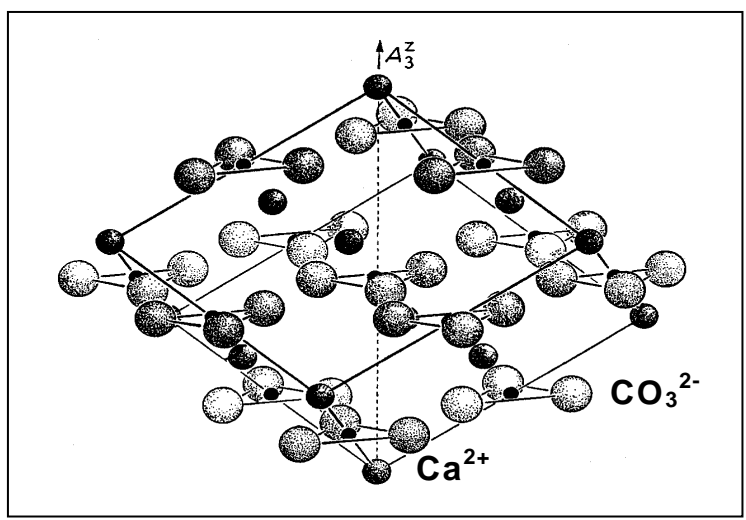

**Abb. 3**: Das trigonale Gitter eines Kalkspat-Rhomboeders (aus HELLWEGE [4]). Die Hauptachse ist gestrichelt gezeichnet.

Bei einer Anregung des Carbonat-Ions, z. B. durch eine Drehung oder Schwingung um die dreizählige Drehachse ist ein anderer Energiebedarf notwendig – also unterschiedliche Eigenfrequenzen - als senkrecht zu dieser Achse. Deshalb existieren zwei verschiedene Brechzahlen. Dass die absorbierte Frequenz in den Wert für die Brechzahl eingeht, habe ich übrigens bei der DPG-Tagung 2001 in Bremen [8] vorgestellt bzw. auch auf der GDCP-Tagung in Dortmund [9]. Jedoch die Bestätigung der Annahme ist noch längst nicht erfolgt.

Übrigens sind doppelbrechende Stoffe in einem bestimmten Frequenzbereich prinzipiell auch dichroitisch, auch wenn die Ausnutzung des Dichroismus nicht unbedingt immer praktikabel sein mag.

Als Grundschwingungen des Carbonat-Ions (s. Abb. 4) existieren auf Grund seiner Struktur vier verschiedene Normalmoden (Beschreibung nach WILLIAMS [18] und Indizierung aus [17]). In der IR-, Raman-, UV-Spektroskopie wird oft als Messgröße die Wellenzahl benutzt, die proportional zur Energie ist (Wellenzahl = Frequenz durch Vakuumlichtgeschwindigkeit). Man unterscheidet grob zwischen Streck- und Biegeschwingungen. Bei einer Streckschwingung verändern sich Bindungsabstände und bei einer Biegeschwingung Bindungswinkel. Es existieren auch Mischformen beider Schwingungstypen. Im folgenden wird auch für die Wellenzahl das gleiche Symbol $\nu$ wie für die Frequenz verwendet.

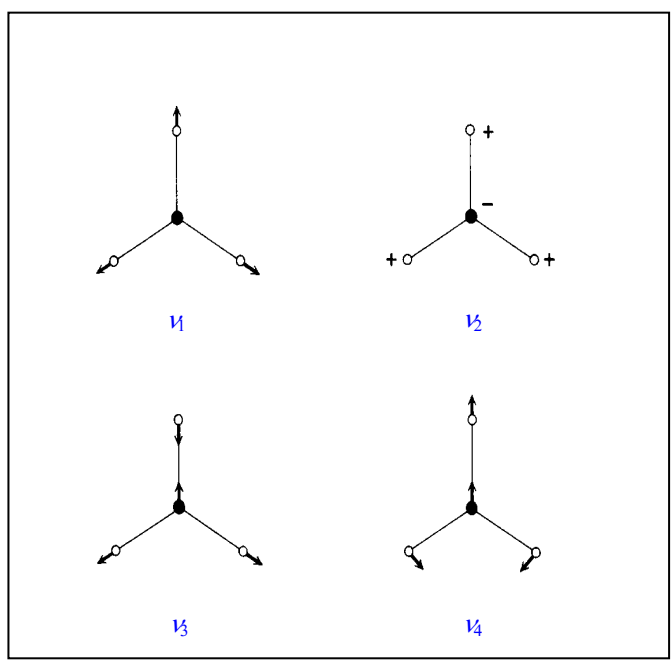

**Abb. 4**: Die vier möglichen Grundschwingungen des Carbonat-Ions (nach [17]). Durch das C-Atom (schwarzer Kreis) verläuft die optische Achse senkrecht zur Papierebene. Die kleinen weißen Kreise entsprechen den O-Atomen. Die Pfeile symbolisieren Teilschwingungen in der Anfangsphase, die durch Photonen bewirkt wurden.

Die Grundschwingungen werden des Carbonat-Ions werden nachstehend in der Reihenfolge von hohen zu niederen Wellenzahlen beschrieben:
1. $\nu_3$ bei 1435 cm$^{-1}$, die asymmetrische Streckschwingung, zweifach entartet,
2. $\nu_1$ bei 1089 cm$^{-1}$, die symmetrische Streckschwingung oder Atmungsschwingung,
3. $\nu_2$ bei 874 cm$^{-1}$, das C-Atom schwingt durch die Ebene der drei O-Atome. Diese Schwingung verläuft in bzw. parallel zur optischen Achse (Parallelschwingung).
4. $\nu_4$ bei 713 cm$^{-1}$, die ebene Biege- und Streckschwingung, zweifach entartet.

Die $\nu_1$ und die beiden entarteten Schwingungen $\nu_3$ und $\nu_4$ verlaufen senkrecht zur optischen Achse, es sind also Senkrechtschwingungen. Im Ramanspektrum sind nur drei Grundschwingungen sichtbar (s. Abb. 5 oben), die $\nu_1$, $\nu_3$ und $\nu_4$. Neben den inneren Schwingungen des Carbonations existieren natürlich noch die äußeren Gitterschwingungen der $Ca^{2+}$- und $CO_3^{2-}$-Ionen, die aber im folgenden nicht weiter benötigt werden. Wer sich nicht für Spektroskopie interessiert, kann schon zum Kapitel 2.2.5 springen.

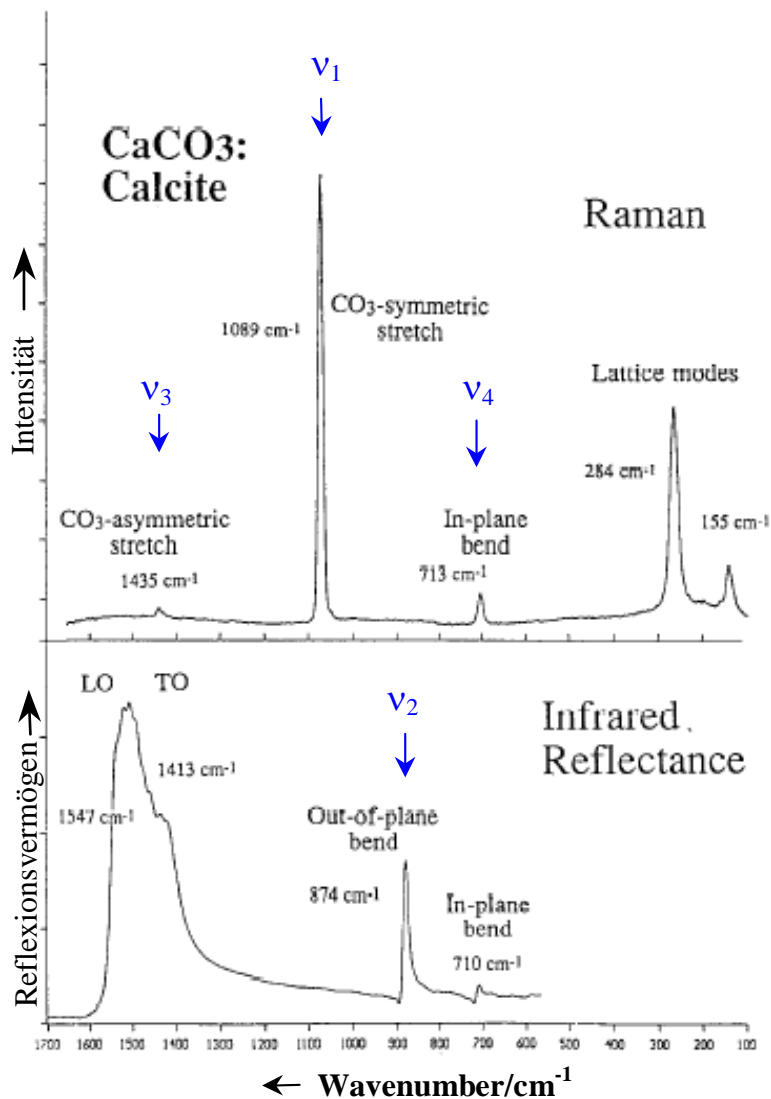

**Abb. 5**: Das Ramanspektrum und Infrarot-Refexionsspektrum von Calcit
(nach WILLIAMS [18])

### 2.2.2 IR-Reflexionsspektren *

Es genügt eigentlich, das reine Ramanspektrum zu betrachten. Der Vollständigkeit halber wurde auch gleich noch das Infrarot-Reflexionsspektrum mit abgebildet, weil es für Insider der Spektroskopie außerdem sehr instruktiv ist. Wie im Ramanspektrum sind auch im Infrarotspektrum nur drei Grundschwingungen zu sehen: die $\nu_3$, $\nu_2$ und $\nu_4$ (Abb. 5 unten sowie Abb. 6c). Anstatt der $\nu_1$ in Ramanspektrum taucht im Infrarotspektrum die $\nu_2$ auf. Die $\nu_3$ ist sehr breit und aufgespalten in einen longitudinal optischen (LO) und einen transversal optischen Peak (TO). Die Bezeichnungen longitudinal und transversal beziehen sich auf die Hauptachse. Während longitudinale optische Photonen in der Absorptionsspektroskopie so gut wie keine Rolle spielen, so sind sie um so bedeutender für die Reflexionsspektroskopie.

Hier noch eine kurze Bemerkung zum Reflexionsvermögen, die im folgenden aber nicht so wichtig ist und nur für den fortgeschrittenen Kenner der Optik gedacht ist. Die Frequenzen können bei schrägem Lichteinfall gegenüber einem Absorptions- oder auch Ramanspektrum (Emissionsspektrum) etwas zu höheren Frequenzen verschoben sein (siehe z. B. HELLWEGE [4], S. 103). Das Reflexionsvermögen $R$ ist definiert als

$$R = \frac{(n-1)^2 + \kappa^2}{(n+1)^2 + \kappa^2} \qquad \text{(Gl. 1)}$$

mit $n$ Brechzahl und $\kappa$ Absorptionskoeffizient. Das nächste Reflexionsspektrum taucht in Abbildung 6 auf.

### 2.2.3 Zur Natur der Photonen allgemein *

Das Thema Photonen kann schier unerschöpflich sein, da man für jedes Teilchen ein spezielles Photon definieren kann, ähnlich wie man für verschiedene Nukleonen unterschiedliche Namen erfunden hat, z. B. Proton für den Kern des Wasserstoffs, Deuteron für den Kern des schweren und Triton für den Kern des überschweren Wasserstoffs. Ganz so schwierig muss man es für das Photon nicht unbedingt gestalten, vor allem, wenn man keinen unübersichtlichen „Photonenzoo" pflegen möchte. Der unüberschaubare „Teilchenzoo" genügt. Viel wichtiger ist ein grundsätzliches Verständnis für das Photon. Aber hierzu gibt es erst mal kleine Hindernisse: Photonen kann man gar nicht direkt sehen wegen ihrer Schnelligkeit, die von keinem Teilchen übertroffen werden kann. Man kann Photonen nur indirekt erkennen an den Änderungen, die sie an Teilchen hervorrufen. Bei der letzten Jahrestagung der DPG in Leipzig habe ich etwas zum Aussehen der Photonen gesagt (SCHULZ 2002 [7]). Danach haben Photonen prinzipiell die gleiche Struktur wie die Teilchen, mit denen sie wechselwirken. Photonen sind das „Negativ" von Teilchen. Denn Photonen haben nämlich gerade das, was die Teilchen vor dem Zusammenstoß (der Wechselwirkung) noch nicht besessen haben. (Übrigens auf den DPG-Tagungen Didaktik der Physik habe ich seit 1997 immer mal kleine Neuigkeiten zum Photon ergänzt, die mir in der Zwischenzeit aufgefallen waren, z. B. [7], [8], [10].)

### 2.2.4 Photonenabkömmlinge *

Hier noch eine weitere letzte kleine Problematik: Selbst wenn man Photonen durch Gegeneinanderschießen in einer Laserapparatur gefangen halten kann, so ist das entstandene Gebilde kein Photon mehr, sondern eher Materie („kondensiertes Licht") oder manchmal speziell photonische Kristalle, im Idealfall ein ruhendes Teilchen mit großer Lebensdauer, ein Bose-Einstein-Kondensat im Ortsraum, das dann also nur so dumm rumliegt und ein reines Teilchen ist.

Die Wechselwirkung zwischen Teilchen und Licht kann interessante angeregte Teilchen hervorbringen, die unter der Bezeichnung Polariton (siehe z. B. WYNANDS [16]) oder Soliton bekannt sind, also Zwitter zwischen Teilchen und Photonen. Schon vor über 50 Jahren wurden Staubpartikel untersucht, die sich bei Lichtbestrahlung auf schraubenförmigen Bahnen bewegten. (Heute würde man solche Teilchen vielleicht als „Staubsolitonen" bezeichnen). Dieses ist ein noch immer nicht verstandenes Kapitel der Physik, bekannt unter dem Schlagwort Photophorese (EHRENHAFT 1951 [2]). Besonders auffallend und instruktiv sind solche angeregten Teilchen, die regelmäßige Bahnen einschlagen.

Auch das moderne Kapitel Solitonen ist im Moment nicht sonderlich aufschlußreich. Aber dennoch, in neuerer Zeit sind schon etliche Solitonen experimentell gesichtet sowie mathematisch und graphisch modelliert worden (WEISS und Mitarbeiter 1995 [14] und 1999 [15]). Die Literaturstelle [14] ist besonders für nicht abstrakt denkende Leser/innen zu empfehlen.

### 2.2.5 Polarisation und Schwingung

Bei der letzten DPG-Tagung 2002 [7] hatte ich am Beispiel eines gasförmigen fünfatomigen Moleküls mit dreizähliger Drehachse wie Methylchlorid die Parallelschwingung (Schwingung in bzw. parallel zur Figurenachse) und die Senkrechtschwingung vorgestellt (Schwingung senkrecht zur Figurenachse). Diese Bezeichnungen gelten natürlich auch für einkristalline Festkörper. Nur heißt hier die Figurenachse eben Hauptachse.

Am Beispiel von $FeCO_3$ (Eisenspat), das isomorph, also baugleich, zu $CaCO_3$ (speziell Calcit) ist, soll an Hand der inneren Schwingungen des propellerförmigen Carbonat-Ions etwas Wichtiges demonstriert werden, das den Einstieg zum Verständnis der linearen Polarisation erleichtert. In Abb. 6 ist ein Infrarotspektrum von $FeCO_3$ dargestellt (ebenfalls als Reflexionsspektrum).

Wenn die Schwingungsrichtung des Lichtstrahls und der Schwingung im Kristall parallel verlaufen, dann ist ein Absorptionspeak bzw. Reflexionspeak durch Reststrahlen im Spektrum beobachtbar. In Abb. 6a ist durch parallel polarisiertes Licht nur die Parallelschwingung $v_2$ und in Abb. 6b sind durch senkrecht polarisiertes Licht nur die beiden Senkrechtschwingungen $v_3$ und $v_4$ zu sehen. Natürliches Licht dagegen (unpolarisierte Strahlung, die Licht aller Polarisationsrichtungen enthält) gibt alle drei Schwingungen zu erkennen (Abb. 6c).

**Also, ein Peak ist nur beobachtbar, wenn für eine vorgegebene Frequenz ein Teilchen da ist, das genauso schwingen kann wie das anfliegende und dann auftreffende Photon, also in derselben Schwingungsrichtung.**

### 3. Einfache Polarisationsversuche mit linear polarisiertem Licht

Ein Polarisator oder Analysator ist ein optischer Baustein, der aus unpolarisiertem Licht eine Schwingungskomponente herausfiltert, und zwar durch Absorption (direkt z. B. durch einen dichroitischen Stoff oder indirekt durch ein Nicolsches Prisma. Das Nicolsche Prisma zerteilt die beiden Lichtkomponenten auf zwei Strahlrichtungen, eine Komponente wird durch ein anderes Material absorbiert).

Wie aus der Abbildung 4 ersichtlich, setzt sich der Gesamtvektor einer Schwingung von drei- und mehratomigen Molekülen oder Ionen aus einzelnen Teilvektoren zusammen. Auf die Betrachtung der Teilvektoren wollen wir im folgenden verzichten und uns nur auf die Resultierende konzentrieren, also auf nur einen Vektorpfeil der hin- und herschwingt. Das Photon, das in den Abbildungen 5 und 6 den Schwingungszustand des Teilchens im Analysator ändert, wird im folgenden kurz als Lichtvektor umschrieben.

Das **Photon der linearen Polarisation**, der Lichtvektor, ist also ein Lichtteilchen, das im einfachsten Fall hin- und her**schwingt**, und sich gleichzeitig senkrecht zur Schwingungsrichtung bewegt (ist also eine **Transversalwelle**), es breitet sich mit Vakuumlichtgeschwindigkeit durch das Vakuum aus (durch Materie etwas langsamer). Sein **Weg-Zeit-Gesetz** ist eine **Sinuskurve**, das entspricht also einer **reellen $\Psi$-Funktion**.

Wir wollen drei Grenzfälle unterscheiden:
- Ein senkrecht schwingendes Photon trifft auf einen Analysator mit gleicher Durchlassrichtung.
- Ein senkrecht schwingendes Photon trifft auf einen gekreuzten Analysator mit waagerechter Durchlassrichtung.
- Ein schräg schwingendes Photon stößt auf einen Analysator mit senkrechter (oder waagerechter) Durchlassrichtung.

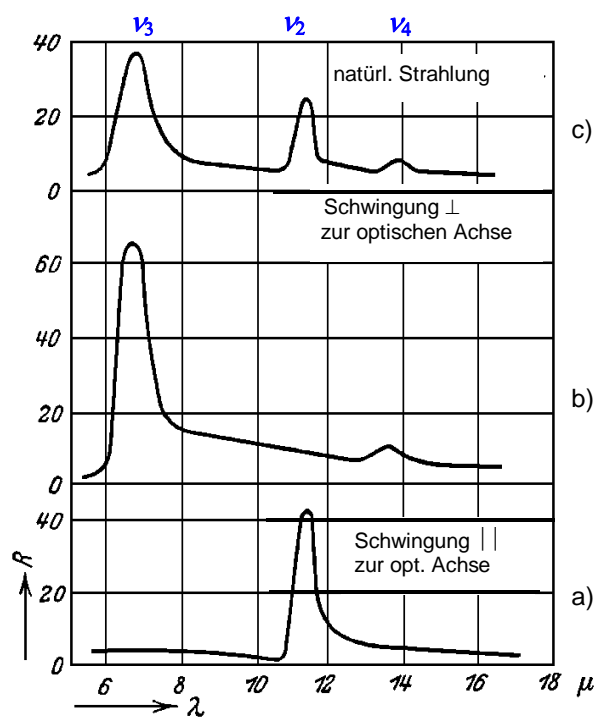

**Abb. 6**: Das Reflexionsvermögen $R$ in % von $FeCO_3$ gegen die Wellenlänge $\lambda$ (nach [4]).
a) parallel polarisiertes, b) senkrecht polarisiertes, c) unpolarisiertes Licht

### a) Senkrechter Lichtvektor in Durchlassrichtung

Stellen wir uns einen Analysator vor mit senkrechter Durchlassrichtung. Trifft ein senkrecht schwingendes Photon (gleich senkrecht schwingender Lichtvektor) also parallel zur Durchlassrichtung des Analysators A, so darf das Licht hindurchgehen. Das Photon mag kurzzeitig absorbiert werden, wird aber sicher durch ein nachfolgendes Photon vertrieben, also emittiert. Prinzipiell kann das gesamte Licht durchgelassen werden (bei Stoffen mit einer Brechzahl gleich 1). Auf alle Fälle wird maximale Helligkeit beobachtet, Abb. 7a.

### b) Waagerechter Lichtvektor (quer zur Durchlassrichtung)

Lichtvektor und Analysatordurchlassrichtung stehen senkrecht. Durch Wechselwirkung des Photons mit der Materie wird es im Analysator absorbiert. Es wird maximale Dunkelheit beobachtet (Abb. 7b).

### c) Schräg schwingender Lichtvektor

Trifft ein schräg schwingender Lichtvektor relativ zur Durchlassrichtung des Analysators auf das Material des Analysators auf, so wird das Photon durch ein schwingungsfähiges und womöglich bereits schwingendes Teilchen in ein senkrecht und in ein waagerecht schwingendes Photon zerlegt: Das waagerecht schwingende Photon wird gemäß b) absorbiert. Dadurch wird die Frequenz automatisch vektorgerecht zerhackt. Das senkrecht schwingende Photon wird gemäß dem Fall a) ausgestoßen. Von der ursprünglich vorhandenen Intensität des Lichtvektors ist nur noch ein Teil durchgelassen (Abb. 8). Den quantitativen Zusammenhang gibt das Gesetz von Malus an, worauf ich nicht eingehen will.

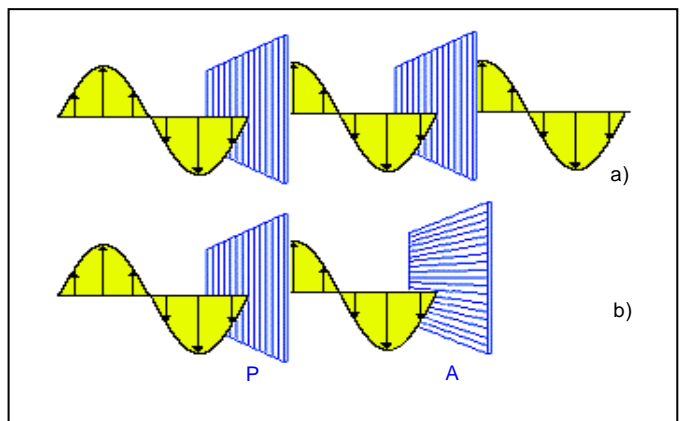

**Abb. 7**: Extremfälle der linearen Polarisation mit einem schwingenden Lichtvektor, dargestellt als gelbe Sinuswelle (Seitenansicht), Abbildung aus [19]
a) Schwingungs-Durchlassrichtung von Analysator A und Polarisator P stehen parallel: der Fall maximaler Helligkeit
b) Analysator- und Polarisatorstellung senkrecht: maximale Dunkelheit

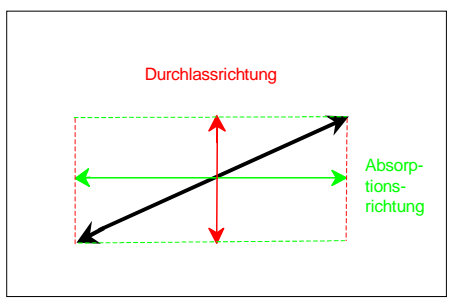

**Abb. 8**:
Der schräg zur Durchlassrichtung des Analysators schwingende Lichtvektor (dicker schwarzer Pfeil) wird vektormäßig aufgeteilt: in eine waagerechte Komponente, die absorbiert wird, und den Rest, also in die senkrechte Komponente, die durchgelassen wird.

Den gleichen Polarisationseffekt wie in den Abbildungen 7 und 8 könnte man mit einer stehenden Seilwelle und einem dazu passenden Gitter aus Lattenrosten erzielen, s. Abb. 9. Das Seil müsste dann hinter dem Analysator befestigt werden.

Die gleichen Vorgänge wie im Analysator haben sich bereits im Polarisator abgespielt. Für den eiligen Leser ist der Artikel hiermit beendet.

### 4. Zum Spin des Photons bei Polarisationsversuchen *

Analog wie es diamagnetische, paramagnetische bis ferromagnetische Stoffe gibt, wird das auch für Photonen gelten. Während die Photonen in der Spektroskopie einen Spin tragen müssen, gilt das nicht zwangsläufig auch für die Lichtteilchen bei Polarisationsversuchen. Wenn eine reine Schwingungsanregung beim Polarisationsexperiment vorliegt, dann sind die anregenden Photonen (Schwingungsphotonen) eben **diamagnetisch**, also spinlos. Es ist zweckmäßig, sich die Schwingungsphotonen aus zwei gegenläufigen Photonen vorzustellen, die mit gleicher Frequenz auf der gleichen Kreisbahn fliegen. Die Kernspin- und Elektronenspinphotonen sind im einfachsten Fall rotierende Kügelchen mit einem Spin, der allerdings nicht um jeden Preis den Wert 1 $\hbar$ tragen muss (s. SCHULZ 2000 [10]), wie es bislang die Macht der Gewohnheit geboten hat.

### 5. Literatur

[1] EHRENHAFT, F.: Über die Photophorese, die wahre magnetische Ladung und die schraubenförmige Bewegung der Materie in Feldern. 1. Teil. In: *Acta Physica Austriaca* 4 (1951), S. 661-488

[2] EHRENHAFT, F.: Über die Photophorese, die wahre magnetische Ladung und die schraubenförmige Bewegung der Materie in Feldern. 2. Teil. In: *Acta Physica Austriaca* 5 (1951), S. 12-29

[3] HECHT, E.: *Optik*, 3. Aufl. München: Oldenburg 2001

[4] HELLWEGE, K.-H.: *Einführung in die Festkörperphysik*. Berlin: Springer 1976

[5] KING, H.; DEKER, U.: Chaos im Laser. In: *Bild der Wissenschaft* 9 (1983), S. 12-14

[6] SCHILLER, S.; MEYN, J.-P.: Von der Photonenspaltung zum kontinuierlich emittierenden Universallaser – Optisch parametrische Oszillatoren ergänzen herkömmliche abstimmbare Laser. In: *Physik-Journal* 1 (2002), S. 35-41

[7] SCHULZ, P.: Infrarotspektroskopie plausibel. In: *CD zur Frühjahrstagung des Fachverbandes Didaktik der Physik in der Deutschen Physikalischen Gesellschaft*, Leipzig 2002. http://arxiv.org/abs/physics/0307062 (Stand: September 2006)
SCHULZ, P.: Infrared Spectroscopy Plausible. In: Progress in Chemical Physics Research. Editor: Linke, A. N., S. 121-135. New York: Nova Science Publishers, Inc. 2005. - ISBN 1-59454-451-4

[8] SCHULZ, P.: Plausible Erklärungshinweise gegen die Überlichtgeschwindigkeit. In: *CD zur Frühjahrstagung Didaktik der Physik in der Deutschen Physikalischen Gesellschaft*, Bremen 2001. ISBN 3-931253-87-2 http://home.arcor.de/gruppederneuen/Seiten/Publikationen/GTunnelT7x.pdf (Stand September 2006)

[9] SCHULZ, P.: Plausible Erklärungshinweise gegen die Überlichtgeschwindigkeit. In: BRECHEL, R. (Herausg.): *Zur Didaktik der Chemie und Physik. Vorträge auf der Tagung für Didaktik der Physik / Chemie in Dortmund, September 2001*. Alsbach: Leuchtturm-Verlag 2001, S. 304-306

[10] SCHULZ, P.: Plausible Definition von Masse, Ladung und Spin. In: *CD zur Frühjahrstagung Didaktik der Physik in der Deutschen Physikalischen Gesellschaft*, Dresden 2000 - ISBN 3-931253-71-6. http://home.arcor.de/gruppederneuen/Seiten/Publikationen/DefinitionMasseLadung.pdf (Stand: September 2006)

[11] WEISS, C. O.; KING, H.: Oscillation Period Doubling Chaos in a Laser. In: *Optics Commun.* 44, Heft 1(1982), S. 59-61

[12] WEISS, C. O.; GODONE, A.; OLAFSSON, A.: Routes to chaotic emission in a cw He-Ne laser. In: *Phys. Rev.* 28A (1983), S. 892-895

[13] WEISS, C. O.; VILASECA, R.: *Dynamics of Lasers*. Weinheim: VCH 1991

[14] WEISS, C. O.; STALIUNAS, K.; VAUPEL, M.: Die rastlosen Wirbel. In: *Physik in unserer Zeit* 26 (1995), S. 169-175

[15] WEISS, C. O.; VAUPEL, M.; STALIUNAS, K.; SLEKYS, G.; TARANENKO, V. B.: Solitons and vortices in lasers. In: *Appl. Phys.* B 88 (1999), S. 151-168
Zusätzliche Videos unter der Web-Adresse http://www.springerlink.com (Stand: September 2006)

[16] WYNANDS, R.: Fortschritt durch Stillstand. In: *Phys. Blätter* 57 (2001), S. 18-20

[17] (bis 2003) http://mailbox.univie.ac.at/werner.mikenda/MS_6.doc

[18] WILLIAMS, Q.: *Mineral Physics and Crystallography. A Handbook of Physical Constants*. AGU Reference Shelf 2. American Geophysical Union 1995, S. 291-302 http://www.agu.org/reference/minphys/18_williams.pdf (Stand: September 2006)

[19] (bis 2002) http://abalone.cwru.edu/tutorial/enhanced/files/lc/light/light.htm

[20] http://www.sue.shiga-u.ac.jp/WWW/dept/butsuri/kyouzai/E.pdf (Stand: Juni 2006)

[21] http://hyperphysics.phy-astr.gsu.edu/hbase/phyopt/polabs.html (Stand: September 2006)

[22] SCHULZ, P.: Photon plausibel bei linearer Polarisation. In: *CD zur Frühjahrstagung des Fachverbandes Didaktik der Physik in der Deutschen Physikalischen Gesellschaft*, Augsburg 2003. – ISBN 3-936427-71-2

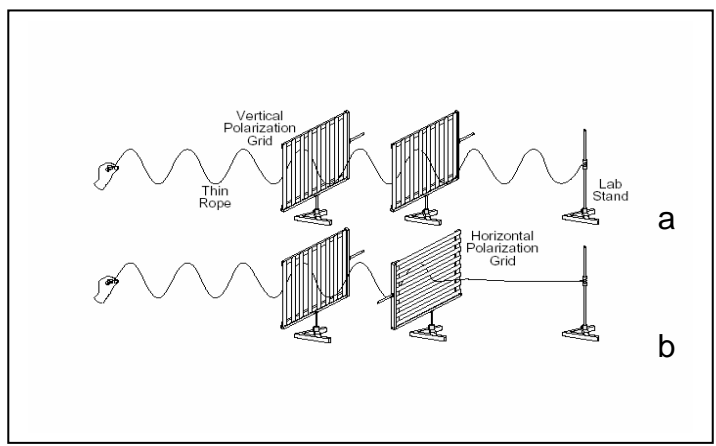

Abb. 9: Polarisationsversuch mit einer Seilwelle an Lattenrosten [20]
    a) parallele senkrechte Polarisation
    b) gekreuzte Lattenroste